\documentclass[aps,prl,superscriptaddress,showpacs,twocolumn]{revtex4-2}
\usepackage[left=1.9cm,right=1.9cm,bottom=1.9cm,top=1.9cm,ignoreall]{geometry}
\usepackage[utf8]{inputenc}
\usepackage{subfigure}
\usepackage{graphicx}
\usepackage[skip=0pt, indent=20pt]{parskip}
\usepackage{amsmath}
\usepackage{amssymb}
\usepackage{xspace}
\usepackage{bm}
\usepackage{color}
\usepackage[squaren,Gray]{SIunits}
\usepackage{empheq}
\usepackage{braket}
\usepackage{physics}
\usepackage{nameref}
\usepackage[hyperindex=true]{hyperref}
\hypersetup{linktocpage,colorlinks=true,citecolor=blue,linkcolor=blue}
\makeatletter
\let\revtex@section\section
\renewcommand{\section}{%
  \@ifstar
    {\@dblarg\revtex@star@section}
    {\@dblarg\revtex@normal@section}}
    
\def\revtex@normal@section[#1]#2{%
  \gdef\@currentlabelname{#2}%
  \revtex@section[#1]{#2}%
}
\def\revtex@star@section[#1]#2{%
  \gdef\@currentlabelname{#2}%
  \revtex@section*{#2}%
}
\makeatother

\begin{document}

\title{Biphoton state generation and engineering with bright hybrid III-V/Silicon photonic devices}

\author{L. Lazzari}
\affiliation{Université Paris Cité, CNRS, Laboratoire Matériaux et Phénomènes Quantiques, 75013 Paris, France}
\affiliation{Université Paris-Saclay, CNRS, Centre de Nanosciences et Nanotechnologies, 91120 Palaiseau, France}
\affiliation{STMicroelectronics, Technology $\And$ Design Platform, 38920 Crolles, France}

\author{J. Schuhmann}
\affiliation{Université Paris Cité, CNRS, Laboratoire Matériaux et Phénomènes Quantiques, 75013 Paris, France}
\affiliation{Université Paris-Saclay, CNRS, Centre de Nanosciences et Nanotechnologies, 91120 Palaiseau, France}
\affiliation{STMicroelectronics, Technology $\And$ Design Platform, 38920 Crolles, France}

\author{O. Meskine}
\affiliation{Université Paris Cité, CNRS, Laboratoire Matériaux et Phénomènes Quantiques, 75013 Paris, France}

\author{M. Morassi}
\affiliation{Université Paris-Saclay, CNRS, Centre de Nanosciences et Nanotechnologies, 91120 Palaiseau, France}

\author{A. Lemaître}
\affiliation{Université Paris-Saclay, CNRS, Centre de Nanosciences et Nanotechnologies, 91120 Palaiseau, France}

\author{M.I. Amanti}
\affiliation{Université Paris Cité, CNRS, Laboratoire Matériaux et Phénomènes Quantiques, 75013 Paris, France}

\author{F. Boeuf}
\affiliation{STMicroelectronics, Technology $\And$ Design Platform, 38920 Crolles, France}

\author{F. Raineri}
\affiliation{Université Côte d’Azur, CNRS, Institut de Physique de Nice, 06200, Nice, France}

\author{F. Baboux}
\affiliation{Université Paris Cité, CNRS, Laboratoire Matériaux et Phénomènes Quantiques, 75013 Paris, France}

\author{S. Ducci}
\thanks{Corresp. author: sara.ducci@u-paris.fr}
\affiliation{Université Paris Cité, CNRS, Laboratoire Matériaux et Phénomènes Quantiques, 75013 Paris, France}

\date{\today}

\begin{abstract}

Hybrid photonic circuits, harnessing the complementary strengths of multiple materials, represent a key resource to enable compact, scalable platforms for quantum technologies. In particular, the availability of bright sources of tunable biphoton states is eagerly awaited to meet the variety of applications currently under development. In this work we demonstrate a heterogeneously integrated device that merges biphoton generation and on-chip quantum state engineering, combining an AlGaAs photon-pair source with a CMOS-compatible silicon-on-insulator (SOI) circuit. Photon pairs are generated in the C telecom band via spontaneous parametric down-conversion and transferred to the SOI chip through a multimode evanescent coupling scheme. This design achieves a pair generation rate above 10$^{6}$ s$^{-1}$mW$^{-1}$ and a coincidence-to-accidental ratio up to 600. Crucially, the coupling design induces strong and predictable transformations of the biphoton joint spectral amplitude, enabling complex quantum state engineering entirely on-chip in a compact device compliant with electrical pumping.

\end{abstract}

\maketitle

\section*{Introduction}
 
Hybrid quantum photonics is an emerging and rapidly advancing field that combines multiple material platforms to exploit their complementary optical properties for quantum information processing \cite{Elshaari20,Labonte24}. In particular, by integrating III-V semiconductors with silicon photonics, hybrid devices can harness the efficient nonlinearities, direct bandgap, and quantum light generation capabilities of III-V materials \cite{Baboux23} while benefiting from the scalability and mature fabrication infrastructure of silicon technologies \cite{Review_Siew,Feng22}. As such, this approach has successfully led to the development of compact, robust, and multifunctional photonic circuits capable of generating, controlling, and routing quantum states of light on a single chip  \cite{Schuhmann24,kues23}. Therefore, hybrid integration offers a compelling route toward scalable quantum technologies, providing improved performance and functionality over monolithic approaches.

Among the desirable properties of quantum light sources, brightness, spectral bandwidth of the generated state, and precise control over the joint spectral amplitude (JSA) stand out as particularly important for fulfilling the requirements of a wide range of applications. In particular, this last capability is essential for a variety of applications, including the control of photon correlations \cite{Lutz14} and exchange statistics \cite{Francesconi21}, as well as the optimization of quantum Fisher information (QFI) to enhance precision in quantum metrology and sensing \cite{Meskine24}. Several methods have been proposed for JSA engineering, including dispersion engineering \cite{Svozilik11,Javid21}, pump shaping \cite{Francesconi20}, and off-chip filtering \cite{Meskine24}. Interestingly, it has been shown that in hybrid structures including an evanescent coupling region, this can be exploited as a spectral filter or transformer, enabling complex manipulation of the biphoton state without external components \cite{Ding22}. Thus, by carefully engineering the coupling region -- including its geometry, refractive index profile, and polarization dependence -- the joint spectral amplitude (JSA) can be precisely tailored, enabling control over the entanglement properties and spectral bandwidth direcly on-chip.

In this work, building on the results presented in \cite{Schuhmann24}, we demonstrate an AlGaAs/SOI hybrid quantum photonic chip that leverages on a multimode evanescent coupling scheme to combine high brightness and control of the JSA of the generated biphoton state. Photon pairs are generated via spontaneous parametric down conversion in an AlGaAs Bragg reflection waveguide and adiabatically transferred to an underlying adhesively bonded SOI waveguide. The measured pair generation rate exceeds 10$^{6}$ s$^{-1}$mW$^{-1}$. We introduce a predictive method to evaluate how the geometry of the coupling region affects the JSA of the produced biphoton state, in both amplitude and phase, thus enabling state engineering at the design stage. The validity of our model is confirmed through Hong-Ou-Mandel (HOM) interferometry \cite{Hong87}. Finally, we demonstrate the versatility of this approach through two representative use cases: the control of photon exchange statistics and the engineering of biphoton states for quantum-enhanced HOM metrology, highlighting its advantages over fully monolithic implementations.

\begin{figure*}[t] % mettere figura con sketch + light intensity e forma della JSA (sezioni 1 e 2 insieme) + index maps
    \includegraphics[width= \textwidth]{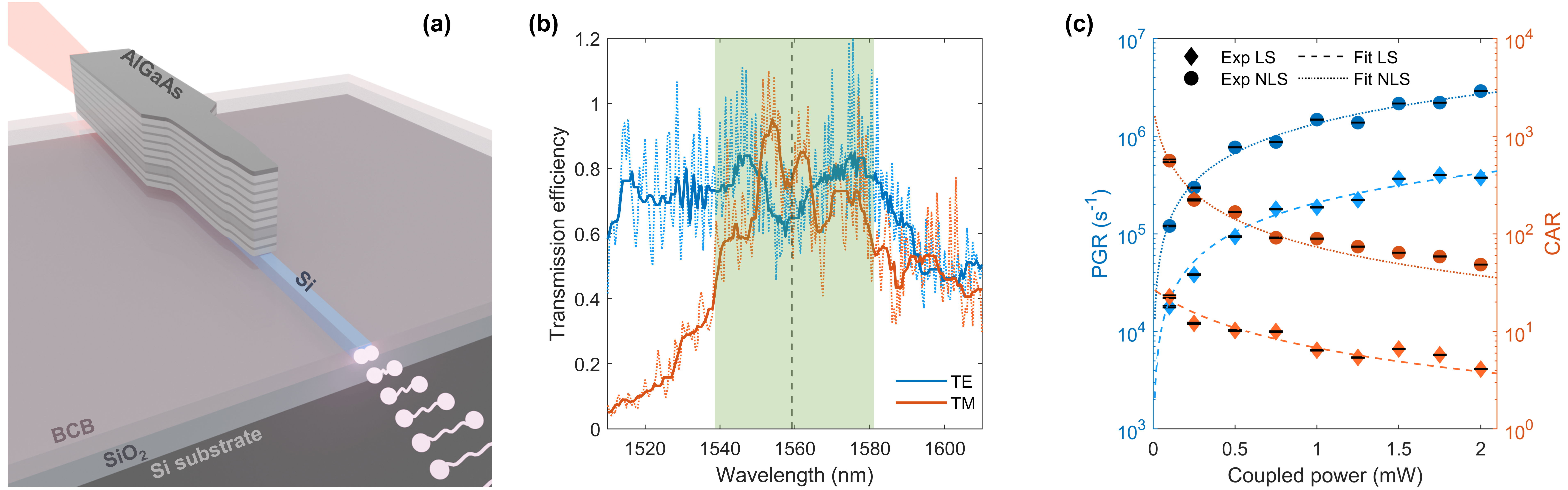}
    \caption{(a) Schematic illustration (not to scale) of the hybrid device and its working principle. (b) Measured transmission efficiency spectra for TE and TM polarizations. Cavity oscillations, visible in the dotted lines, are averaged out in the solid lines. The green-shaded area marks the spectral region around the SPDC resonance (indicated by the vertical grey line) where the transmission efficiency exceeds 50\%. (c) Internal PGR and CAR as functions of the coupled pump power, for the linearly shaped (LS, from \cite{Schuhmann24}) and nonlinearly shaped (NLS, this work) tapers. Curve fittings are included to guide the eye and illustrate the expected trends; error bars are estimated assuming Poissonian statistics for the coincidence counts.}
     \label{fig:device}
\end{figure*}

\section*{Photon-pair generation}

The AlGaAs source is a nano-fabricated Bragg reflection waveguide (BRW), 2 mm long and 4 µm wide (see \nameref{sec:methods}), designed to generate photon pairs via spontaneous parametric down-conversion (SPDC) at telecom wavelengths and room temperature \cite{Review_Feli}. Heterogeneous integration with silicon is achieved through adhesive bonding \cite{Crosnier17} of the III–V stack onto a pre-fabricated 600 nm-thick SOI photonic circuit \cite{Schuhmann24}. After chemically removing the GaAs substrate, the AlGaAs waveguides are patterned and aligned to the underlying silicon structures by electron-beam lithography, and subsequently etched using inductively coupled plasma reactive-ion etching (ICP-RIE), together with the tapered sections required for efficient optical coupling. The facets at both ends of the device are defined by cleaving the sample to the desired length. A schematic of the hybrid device is shown in Fig. \ref{fig:device}a.

The coupling section incorporates a nonlinearly shaped (NLS) taper, designed according to a multimode generalization of the adiabatic criterion presented in \cite{sun2009adiabaticity}. The taper is implemented at the end of the AlGaAs photon-pair generation region, while the underlying silicon waveguide keeps a constant width. This geometry maximizes the coupling efficiency while minimizing the device footprint and power scattering into higher order modes for both TE and TM-polarized fundamental modes in the C telecom band. This is a critical requirement for the efficient transfer of photons generated through type 2 nonlinear conversion process, where a TE-polarized pump generates orthogonally polarized signal and idler photons. The polarization orthogonality enables deterministic photon separation, facilitating operations such as density matrix reconstruction and Hong Ou Mandel (HOM) interferometry \cite{James01}. The measured transmission spectra for an 800 $\mu$m-long taper are shown in Fig. \ref{fig:device}b, exhibiting transmission efficiencies of approximately 80\% for both polarizations and a bandwidth of around 45 nm, well centered at the SPDC degeneracy wavelength (1560 nm).  

\begin{figure*}[t]
    \centering
    \includegraphics[width=\textwidth]{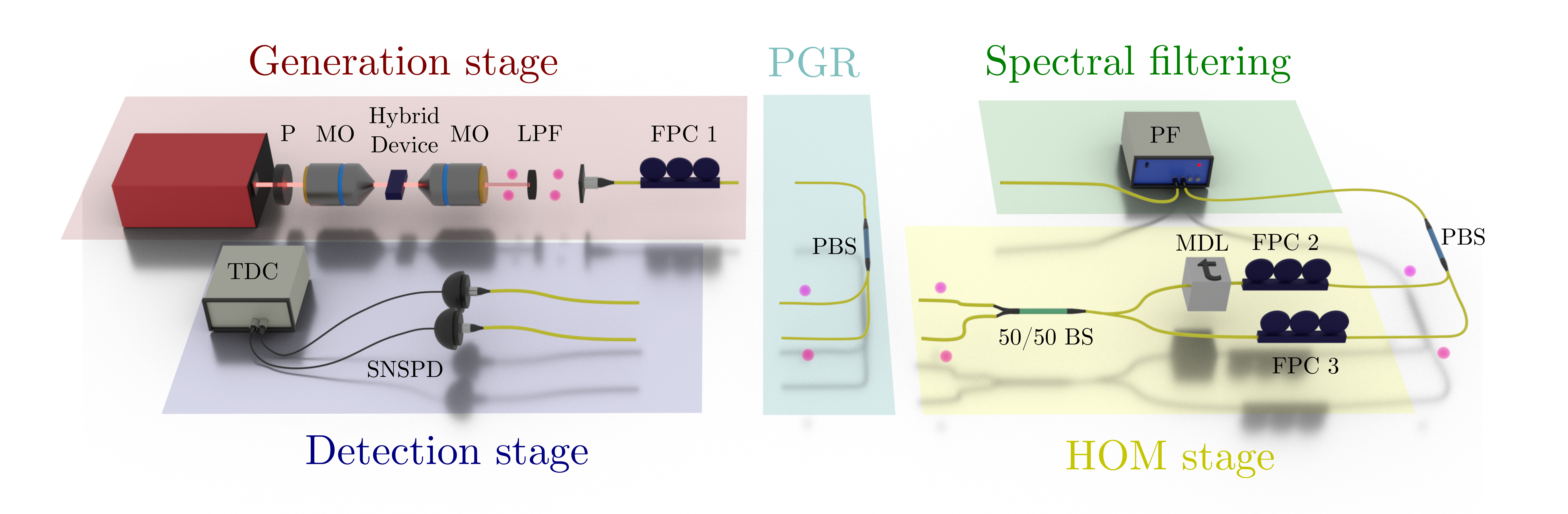}
    \caption{Schematic of the experimental setup. After photon-pair generation and collection, the photons can be directed either to the PGR/CAR/Detection stage, %for pair generation rate (PGR) and coincidence-to-accidental ratio (CAR) measurements
   or alternatively to the HOM stage -- optionally including spectral filtering -- followed by the Detection stage for Hong–Ou–Mandel (HOM) interferometry measurements. P: linear polarizer; MO: microscope objective; WG: waveguide; LPF: long-pass filter; PF: programmable filter}
    \label{fig:setup}
\end{figure*}

To assess the impact of our design on photon-pair generation, we couple a continuous-wave laser beam (TOPTICA, linewidth$\approx$100 kHz) tuned to half the degeneracy wavelength (780 nm) into the AlGaAs waveguide using a microscope objective, as sketched in Fig.\ref{fig:setup}). The generated type 2 photon pairs are collected from the silicon waveguide through another microscope objective, coupled to a SMF-28 optical fiber via a fiber collimator, deterministically separated by a polarizing beam splitter (PBS), and directed to superconducting nanowire single-photon detectors (SNSPDs, Quantum Opus). A fibered polarization controller (FPC1) is used to optimize the PBS splitting ratio. The detection signals are processed by a time-to-digital converter (TDC, quTools). From the coincidence histogram recorded at the detectors, we extract the internal pair generation rate (PGR) -- defined as the rate of photon pairs generated in the AlGaAs waveguide and transferred to the Si waveguide -- and the coincidence-to-accidental ratio (CAR) as functions of the coupled pump power. The results, shown in Fig. \ref{fig:device}c, reveal an order-of-magnitude improvement in both PGR and CAR compared to the linearly shaped (LS) taper configuration reported in Ref. \cite{Schuhmann24}. In particular, the measured PGR exceeds $10^6$ s$^{-1}$ per mW of coupled pump power, while the CAR reaches up to $6\times10^2$ at low pump powers. To the best of our knowledge, these represent the highest values reported to date for hybrid III–V/silicon photon-pair sources.

\section*{Joint spectral amplitude}

We now introduce a predictive model to simulate the influence of the coupling region on the joint spectral amplitude (JSA) of the generated biphoton state. We begin by considering a straight AlGaAs waveguide without optical coupling to silicon, pumped by a continuous-wave (CW) monochromatic laser. In this case, the biphoton state generated via spontaneous parametric down-conversion (SPDC) can be expressed as \cite{giorgio20}:
\begin{equation}
    \ket\psi=\iint \text{d}\omega_s\text{d}\omega_i\phi(\omega_s,\omega_i)\hat{a}_s^\dag(\omega_s)\hat{a}_i^\dag(\omega_i)\ket0, \label{eq:stateJSA}
\end{equation}
where $\hat{a}_x^\dag(\omega_x)$ denotes the creation operator for a photon in mode $x$ with frequency $\omega_x$ (with $x=s,i$) and $\ket0$ represents the vacuum state. The function $\phi(\omega_s,\omega_i)$ is the joint spectral amplitude (JSA): its modulus squared, the joint spectral intensity (JSI), gives the probability of generating a signal photon at frequency $\omega_s$ together with an idler photon at frequency $\omega_i$. The JSA depends on both the phase-matching function $\phi_{PM}$ and the spectral profile of the pump beam $\phi_p$ \cite{Gianani20}, such that $\phi(\omega_s,\omega_i)\propto\phi_p(\omega_s+\omega_i)\phi_{PM}(\omega_s,\omega_i)$. For a sufficiently narrowband pump, typical of continuous wave (CW) operation, $\phi_p$ can be approximated by a Dirac delta function, $\delta(\omega_p-\omega_s-\omega_i)$, where $\omega_p$ is the pump frequency. $\phi_{PM}$ reflects the phase-matching condition and is therefore determined by the physical properties of the device -- specifically, the waveguide geometry, the material's nonlinear response, and its dispersion characteristics, including group velocity dispersion -- and can be expressed as \cite{Yang08}:
\begin{equation}
    \phi_{PM}(\omega_s,\omega_i)\propto\frac{e^{i\Delta k\frac{L}{2}}}{\sqrt{v_g^s(\omega_s)v_g^i(\omega_i)}}\text{sinc}\left(\Delta k\frac{L}{2}\right),
    \label{eq:phi_PM}
\end{equation}
where $\Delta k=k_p-k_s-k_i$ is the phase mismatch, $k_x$ is the wavevector of mode $x$ (signal or idler), $L$ the length of the waveguide, and $v_g^x$ the group velocity of the mode $x$. 
Assuming a perfectly monochromatic pump, both $\phi_{PM}$ and the JSA depends only on the signal frequency, since energy conservation imposes $\omega_i=\omega_p-\omega_s$ with $\omega_p\simeq\text{const}$. Consequently, the JSA reduces to a single-frequency dependence $\phi(\omega_s,\omega_i)\rightarrow\phi(\omega_s)$.

\begin{figure*}[t]
    \centering
    \includegraphics[width=\textwidth]{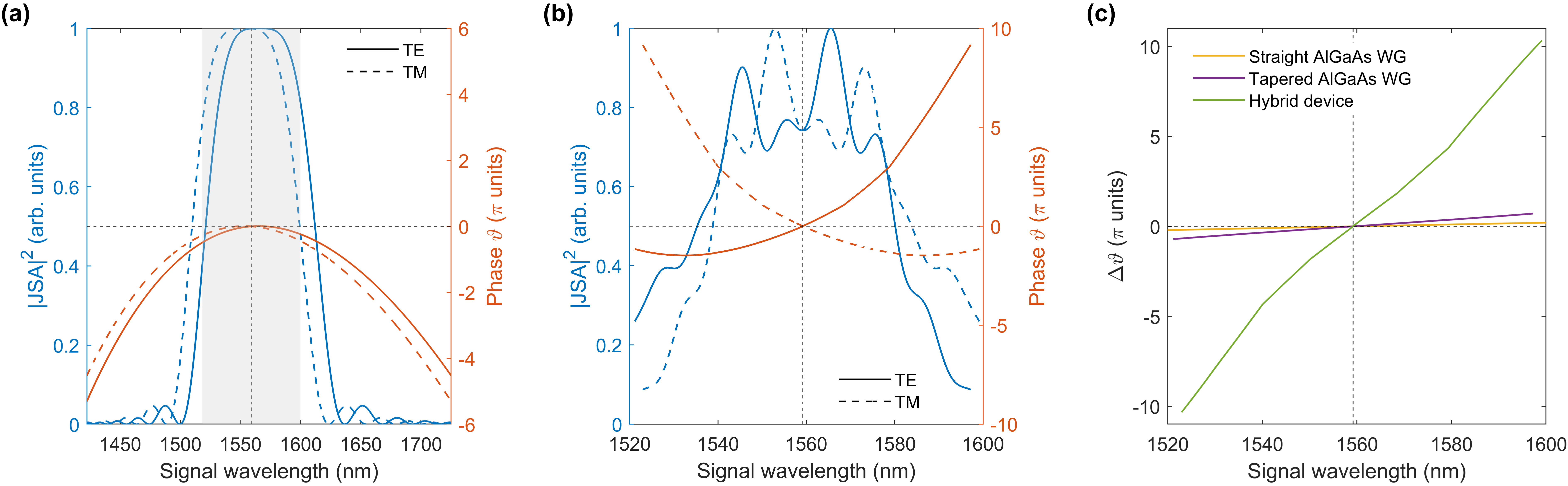}
    \caption{(a) Simulated amplitude and phase of the JSA of the biphoton state generated by a 2 mm-long, 4 $\mu$m-wide AlGaAs waveguide (same parameters as the generation region of the hybrid device).(b) Corresponding simulation for the same AlGaAs waveguide coupled to a 560 nm-wide silicon waveguide via an 800 $\mu$m-long nonlinearly shaped taper. (c) Phase mismatch, $\Delta\vartheta=\vartheta(\omega_s)-\vartheta(\omega_i)$, for a straight AlGaAs waveguide (WG), a tapered AlGaAs WG, and  the hybrid device. The grey-shaded region in (a) indicates the spectral range of interest in (b) and (c).}
    \label{fig:JSA}
\end{figure*}

The JSA is a complex-valued function characterized by both amplitude and phase. When the material parameters are known, it can be numerically simulated. Fig. \ref{fig:JSA}a shows the JSA, $\phi_{BRW}(\omega_s)$, computed using finite-difference eigenmode (FDE) simulations (Lumerical MODE) to extract the modal dispersion of the AlGaAs waveguide, assuming a pump wavelength of 780 nm. The two curves correspond to the two possible polarizations of the generated photons and are symmetric with respect to the degeneracy frequency ($\omega_p/2$). The slight asymmetry arises from the weak birefringence of the waveguide \cite{Appas21}.

When moving to the hybrid device, the joint spectral amplitude (JSA) is modified by the coupling process, affecting both its amplitude and phase. If the joint spectral intensity (JSI) of the photons generated in the AlGaAs waveguide is given by $\left|\phi_{BRW}(\omega_s)\right|^2$ 
then, after mode transfer, the marginal JSI $\left|\phi(\omega_s)\right|^2$ can be expressed as:
\begin{equation}
    \left|\phi(\omega_s)\right|^2=\left|\phi_{BRW}(\omega_s)\right|^2\cdot T_u(\omega_s)\cdot T_v(\omega_p-\omega_s),
    \label{eq:marginal_JSI}
\end{equation}
where $T(\omega)$ denotes the device intensity transmission at frequency $\omega$, $u$ and $v$ represent the mode polarizations, and we have used the relation $\omega_i = \omega_p - \omega_s$. The transmission functions $T_{u,v}(\omega)$ used in this work correspond to the experimentally measured data (Fig. \ref{fig:device}b), allowing a direct link to the spectral response of the specific device. Regarding the phase, the coupling stage introduces a frequency-dependent phase shift $\vartheta_{u,v}(\omega)$, which can be written as:
\begin{equation}
    \vartheta_{u,v}(\omega)=\frac{\omega}{c}\int_0^l n^{u,v}_{eff}(z,\omega)\text{d}z,
    \label{eq:phase_shift}
\end{equation}
where $c$ is the speed of light in vacuum, $l$ is the taper length and $z$ denotes the propagation direction. The effective index profiles $n^{u,v}_{eff}(z,\omega)$ are obtained from finite-difference eigenmode (FDE) simulations, by tracking the evolution of the spatial distribution of the even and odd hybrid modes during the power transfer process \cite{Coupled_Mode_Theory}, at a given operating temperature. The total phase contribution is then given by
$\vartheta(\omega_s)=\vartheta_{u}(\omega_s)+\vartheta_{v}(\omega_p-\omega_s)$.

In Fig. \ref{fig:JSA}b, we present the simulated joint spectral amplitude (JSA) of one of our hybrid devices as a function of the signal wavelength. The corresponding phase mismatch, defined as $\Delta\vartheta=\vartheta(\omega_s)-\vartheta(\omega_i)$ and shown in Fig. \ref{fig:JSA}c, is particularly relevant for HOM interferometry, as will be discussed in the next Section). For comparison, we also report the same quantity for a taper realized solely in the AlGaAs waveguide, i.e., without coupling to the silicon section. In this case, the magnitude of the phase shift is considerably smaller, indicating that efficient phase modulation originates from the coupling process itself. This behavior is a distinctive feature of hybrid devices that integrate dissimilar material platforms, whose contrasting dispersion properties enable enhanced spectral and phase control.

It is worth noting that, throughout this discussion, the cavity oscillations observed in the transmission spectra -- arising from reflections at the device facets -- and their corresponding influence on the JSA \cite{giorgio20} have been averaged out, for simplicity and to ensure consistency with the experimental measurements.

\begin{figure*}[t]
    \centering
    \includegraphics[width=\textwidth]{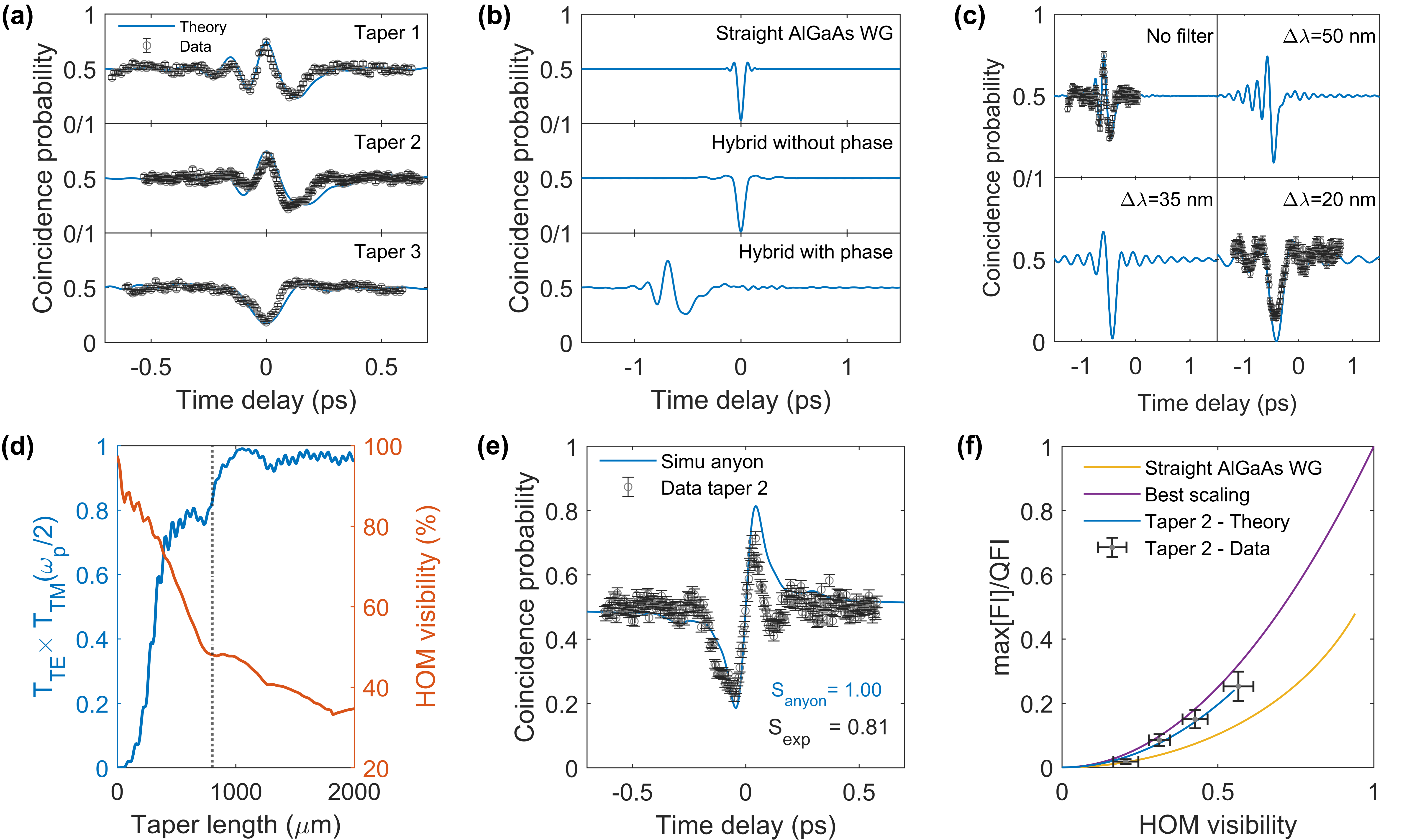}
    \caption{(a) Simulated HOM interferograms (solid lines) and measured data points for hybrid devices with tapers 1, 2, and 3. (b) Top: Simulated HOM interferogram considering the JSA of the biphoton state generated in a 2 mm-long, 4 $\mu$m-wide AlGaAs waveguide. Center: Corresponding simulation for the hybrid device with taper 1, considering only the amplitude modification of the JSA introduced by the coupling region. Bottom: The same simulation including both amplitude and phase modifications. (c) Simulated HOM interferograms (solid lines) for the hybrid device with taper 1 when the JSA spectral width is modified using a rectangular band-pass filter. Experimental data (grey circles) are overlaid for the unfiltered case and for a 20 nm-wide filter. (d) Evolution of the HOM visibility -- assuming unchanged transmission spectra -- and simulated cross transmission ($T_{TE}(\omega_p/2)\times T_{TM}(\omega_p/2)$) as a function of the taper length for the hybrid device with taper 2. The dotted line marks the operating point of the device under study. (e) Simulated HOM interferogram for a phase profile $\vartheta(\omega_s)=\text{sign}(\omega_p-2\omega_s)i\pi/4$, corresponding to $\alpha=1/2$ anyonic particles. Experimental data for the hybrid device with taper 2 are superimposed for  qualitative comparisons. The asymmetry score for both curves is reported. (f) Scaling of the Fisher information (FI) relative to the Quantum Fisher Information(QFI) as function of the HOM visibility for the hybrid device with taper 2. this is compared with the scaling obtained with the full state generated by a straight AlGaAs waveguide and with the best achievable scaling limit. 
    Error bars, where present, are calculated assuming a Poissonian statistics for the coincidences counts.}
    \label{fig:HOM}
\end{figure*}

\section*{Hong-Ou-Mandel interferometry}

We now turn to the experimental validation of the predictive model for biphoton state transformation through the coupling stage discussed in the previous section. To achieve this, a powerful and sensitive tool for experimentally reconstructing the biphoton JSA is HOM interferometry \cite{Tischler15}, and its generalizations \cite{Li24}. In the simpler configuration \cite{Bouchard20}, the two photons are directed to a 50:50 beam splitter (BS), and the coincidence rate is measured as a function of the relative delay $\tau$ between the two photons, which is controlled by introducing a phase shift in one of the interferometer arms. In this way, the HOM interferogram is directly linked to the specific JSA, and the coincidence probability -- assuming no losses and a perfectly balanced (P)BS -- can be expressed as \cite{Descamps23b}:
\begin{equation}
    \begin{aligned}
        P_c&(\tau)= \\
        \frac{1}{2}&\left(1-\text{Re}\iint \text{d}\omega_s\text{d}\omega_i\phi(\omega_s,\omega_i)\phi^*(\omega_i,\omega_s)e^{i(\omega_i-\omega_s)\tau}\right),
   \end{aligned}
    \label{eq:HOM}
\end{equation}
or equivalently, as a function of the signal frequency alone, under the monochromatic pump approximation: $P_c(\tau)=\frac{1}{2}\left(1-\text{Re}\int \text{d}\omega_s\phi(\omega_s)\phi^*(\omega_p-\omega_s)e^{i(\omega_p-2\omega_s)\tau}\right)$, where $\phi(\omega)=\sqrt{|\phi(\omega)|^2}e^{i\vartheta(\omega)}$. Note that, since the JSA function of our device (Fig. \ref{fig:JSA}a,b) is asymmetric with respect to $\omega_p/2$, we have $\phi(\omega_s)\neq \phi(\omega_p-\omega_s)\leftrightarrow \phi(\omega_s)\phi^*(\omega_p-\omega_s)\neq\left|\phi(\omega_s)\right|^2$. As a result, the phase contributions do not cancel out (as shown in Fig. \ref{fig:JSA}c) and directly influence the HOM interference pattern \cite{Lei24,Francesconi21}. 
In particular, expanding $\vartheta(\omega_s)$ to the third order around $\omega_p/2$ yields: $\vartheta(\omega_s)\simeq\vartheta_0+\beta^{(1)}(\omega_p/2-\omega_s)+\beta^{(2)}(\omega_p/2-\omega_s)^2+\beta^{(3)}(\omega_p/2-\omega_s)^3$. The HOM coincidence probability, as expressed in Equation \ref{eq:HOM}, depends on the phase difference $\Delta\vartheta=\vartheta(\omega_s)-\vartheta(\omega_i)$. Due to perfect frequency anti-correlation in the case of a monochromatic pump, we have $\omega_p/2-\omega_s=-(\omega_p/2-\omega_i)$, leading to : $\Delta\vartheta\simeq2\left(\beta^{(1)}(\omega_p/2-\omega_s)+\beta^{(3)}(\omega_p/2-\omega_s)^3\right)$. Thus, only the odd-order terms contribute to the HOM interferogram. The linear term $\beta^{(1)}$ merely shifts the position of the dip, while the cubic term $\beta^{(3)}$introduces a non-trivial modification to the HOM interference pattern \cite{Mazzotta16}.

The HOM setup used for this analysis is depicted in Fig. \ref{fig:setup}). After being separated by the fibered PBS, the photons are directed into the two arms of the HOM interferometer. The polarization is adjusted by two fibered polarization controllers (FPC1 and FPC2), one in each arm, while the temporal delay between the photons is adjusted using a motorized optical delay line (MDL). The two paths are then recombined and separated by a 50:50 beam splitter. Finally, the photons are detected by superconducting nanowire single-photon detectors (SNSPDs), and the temporal correlation histogram is generated by the time-to-digital converter (TDC).

The results of the experimental measurements, overlaid with the simulated curves obtained from the JSA calculated as shown in Fig. \ref{fig:JSA}, are presented in Fig. \ref{fig:HOM}a. The agreement between the experimental data and the theoretical predictions is confirmed for three hybrid devices including tapers with distinct geometrical characteristics (same taper shape in AlGaAs, with silicon waveguide widths of 550, 560 and 570 nm for tapers 1,2, and 3 respectively), and corresponding effective index profiles. This validates our approach for predicting the JSA engineering introduced by the coupling stage. By simulating the (non physical) scenario where the phase shift for the device with taper 1 is neglected, we observe (Fig. \ref{fig:HOM}b) that this phase shift is the primary contribution to the modification of the HOM interferogram compared to the case without coupling stage. Specifically, we expect a shift in the dip position, governed by the linear term $\beta^{(1)}$, and a significant distortion of the dip shape, primarily due to the cubic term $\beta^{(3)}$. The dip shift $\delta\tau_{dip}$ is measured by comparing the absolute dip position in the delay line reference frame for both the hybrid device and the AlGaAs waveguide alone, which are present in the same chip, thus enabling the measurement under practically identical experimental conditions. This provides direct access to the linear dispersion \cite{Steinberg92}, yielding values that closely match our theoretical predictions : $\delta\tau_{dip}=0.52$ ps according to simulations, and $\delta\tau_{dip}=0.50\pm0.01$ ps experimentally for the hybrid device with taper 1. Note that in Fig. \ref{fig:HOM}a,e the experimental curves have been  arbitrarily shifted along the x-axis and the zero delay redefined for clarity.

The cubic dispersion contribution, in contrast, is more challenging to access directly. However, its influence on the HOM interferogram can be investigated by narrowing the spectral width of the JSA, for example, by applying a band-pass filter (BPF) centered at $\omega_p/2$ (Spectral filtering stage in Fig. \ref{fig:setup}). In this case, the contribution of the cubic term is significantly reduced, as we enter a regime where $\beta^{(3)}\left(\omega_p/2-\omega_s\right)^3\ll\beta^{(1)}\left(\omega_p/2-\omega_s\right)$ within the filter bandwidth. Figure \ref{fig:HOM}c shows the simulated HOM interferograms for taper 2 with varying spectral widths of the JSA, filtered through a rectangular BPF. While the symmetric HOM dip \cite{Hong87} is recovered for a filter width of approximately 30 nm -- as experimentally confirmed with a 20 nm-wide BPF -- the dip shift remains virtually unchanged.
The same effect can be achieved by appropriately designing the device, such as by narrowing the transmission bandwidth. This is exemplified by the hybrid device with taper 3 (Fig. \ref{fig:HOM}a), where the phase-related modulation is significantly reduced. Nevertheless, the phase profile still leads to a notable decrease in the dip visibility -- defined here as $V=\left[P_c(\infty) - P_c(\tau-\delta\tau_{dip})\right]/P_c(\infty)$ -- compared to the (quasi-)flat phase case. This reduction is linked to an increased distinguishability between the two photons \cite{Bouchard20}, arising from the frequency dependence of the phase shift. Figure \ref{fig:HOM}d presents the simulated evolution of HOM visibility and crossed transmission efficiency -- defined as $T_{TE}(\omega_p/2)\times T_{TM}(\omega_p/2)$ -- as a function of the taper length for the hybrid device with taper 1. This simulation assumes the (non-physical) scenario where the transmission spectral profile remains the same as that measured for the 800 $\mu$m-long taper (Fig. \ref{fig:device}b) across all taper lengths. Shorter tapers yields higher visibility but at the cost of lower transmission efficiency. Conversely, 
longer tapers, which introduce greater phase shifts, show reduced visibility but enhanced modulation and improved transmission efficiency. Therefore, depending on the specific application, an optimal operating point for the device can be identified. 

We next discuss two representative applications enabled by on-chip JSA engineering. As a first example, we consider the emulation of exchange statistics -- that is, the ability to reproduce the quantum statistical behavior of different classes of particles, such as bosons, fermions, or anyons, by tailoring the symmetry of the biphoton wavefunction \cite{Francesconi20,Francesconi21}. By precisely shaping the phase profile of the JSA through the coupling design, our hybrid platform enables controlled modification of the exchange symmetry of the generated photon pairs. In Fig. \ref{fig:HOM}e, we present a qualitative comparison between the HOM interferogram measured using taper 2 and the corresponding numerical simulation, obtained by applying to the same JSA amplitude a phase profile $\vartheta(\omega_s)=\text{sign}(\omega_p-2\omega_s)\pi/4$. This phase corresponds to the condition required to emulate anyonic particles with exchange statistics $\alpha=1/2$ \cite{Francesconi21}, which are characterized by a perfectly antisymmetric HOM response with respect to the point of coordinates $(0, 1/2)$. To quantify the degree of asymmetry, we define an asymmetry score $S=\sqrt{\int\frac{1}{4}\left[P_c(\tau)-P_c(-\tau)\right]^2\text{d}\tau}/\sqrt{\int\left[P_c(\tau)\right]^2\text{d}\tau}$, which equals 0 for a perfectly symmetric curve and 1 for a perfectly antisymmetric one. The simulated curve yields $S_{anyon}=1.00$, as expected, whereas the experimental value is $S_{exp}=0.81$. The residual discrepancy arises from the fact that the phase profile introduced by the coupling stage does not exactly reproduce the ideal anyonic phase. Nonetheless, the coupling scheme enables a substantial modification of the exchange symmetry -- given that the as-generated biphoton state has $S\simeq0$ --,without requiring any external components or off-chip phase manipulation. With further optimization of the coupling design, phase profiles even closer to the target can be achieved; for instance, simulations predict an increased asymmetry score of $S=0.89$ for a 500 $\mu$m-long taper, requiring only minor adjustments to the present device.

Finally, we explore the potential of our hybrid device for enhanced quantum metrology. HOM interferometry can be harnessed to approach the ultimate quantum precision limit in time-delay estimation \cite{Descamps23b}, enabling attosecond-level resolution \cite{Jordan22}. This limit is attained for states exhibiting perfect HOM visibility ($V=1$) and is fundamentally bounded by the number of experimental repetitions and by the quantum Fisher information (QFI), which depends solely on the JSA amplitude of the probe state \cite{Giovannetti06}. Specifically, the QFI is given by $\text{QFI}=4\Delta^2\omega_s$, where $\Delta^2$ denotes the frequency variance operator. More generally, for states with non-ideal visibility, the achievable precision is determined by the Fisher information (FI), which quantifies the extractable information about a parameter for a given measurement configuration. The QFI thus represents the maximum FI attainable over all possible measurement strategies on the state. The FI can be directly linked to the HOM interferogram through the relation \cite{Descamps23b}: $\text{FI}(\tau)=\left[\partial P_c(\tau)/\partial\tau\right]^2/\left[P_c(\tau)(1-P_c(\tau))\right]$. Consequently, the phase profile of the JSA -- which, as discussed, governs the shape of the HOM interference pattern -- also determines the FI. In the presence of non-ideal visibilities, an important figure of merit for a given state is the scaling of the ratio $\text{max}\left[\text{FI}(\tau)\right]/\text{QFI}$ as a function of $V$. This quantity indicates how closely the achievable precision approaches the quantum limit and provides a measure of the probe state’s robustness within a specific measurement configuration. In Fig. \ref{fig:HOM}g, we present the simulated and measured scaling for the biphoton state generated with the hybrid device with taper 2, compared with that of a straight AlGaAs waveguide and with the theoretical optimal scaling ($\text{max}\left[\text{FI}(\tau)\right]/\text{QFI}=V^2$ \cite{Meskine24}). The introduced phase modulation markedly enhances the scaling relative to the state generated by the straight waveguide, while only slightly reducing the QFI, indicating that a substantial fraction of the quantum limit can be retained even for moderate visibilities. Depending on the target application, the coupling design can thus be tailored to achieve the desired balance between transmission efficiency, HOM visibility -- and consequently $\text{max}\left[\text{FI}(\tau)\right]$ -- and FI scaling. These results demonstrate that our hybrid platform, offering on-chip control over both the amplitude and phase of the JSA, represents a powerful resource for advanced quantum metrology.

\section*{Summary and Conclusion}

In summary, we have demonstrated a hybrid III–V/silicon nonlinear photon-pair source that combines high coupling efficiency over a 40 nm bandwidth for both TE and TM polarizations establishing a new state-of-the-art for the generation rate, bridging the gap between hybrid and monolithic semiconductor platforms. The evanescent coupling between materials with distinct dispersion enables precise amplitude and phase control of the biphoton joint spectral amplitude, allowing on-chip tailoring of quantum correlations, as confirmed by HOM interferometry and theoretical modeling.

Crucially, the influence of the coupling design on the JSA can be accurately predicted and optimized at the device design stage, bringing functionalities previously achievable only with complex off-chip systems into a compact on-chip platform. As demonstrated, the coupling-induced phase profile can be exploited to manipulate photon exchange statistics -- opening prospects for on-chip emulation of anyonic particles with non-trivial statistics \cite{Francesconi21} -- and to enhance quantum metrological performance through engineered biphoton states \cite{Meskine24}. 

Looking ahead, refinements in coupling design and fabrication are expected to further boost transmission efficiency and polarization insensitivity. Numerical estimates indicate that reducing the number of AlGaAs Bragg mirrors on the SOI side from two to one could significantly enhance the coupling strength for both polarizations, yielding a transmission efficiency approaching 0.9 on a bandwidth of 50 nm while halving the coupling length from 500 to 250 $\mu$m.

Regarding phase engineering of the JSA for tailored quantum information applications, an appealing direction lies in inverse design. For instance, one could enforce local adiabaticity while adjusting the taper profile to achieve a desired biphoton phase distribution. In this context, numerical approaches based on differentiable mode solvers offer a highly efficient framework for optimizing waveguide dispersion in broadband phase-matched nonlinear processes \cite{Gray:24}. 

We note that the broad and continuous emission bandwidth of our hybrid device enables frequency-entangled photon generation for high-dimensional \cite{Cozzolino2019, kues23} and continuous-variable quantum information protocols \cite{DescampsPRL23CV}. Combined with polarization versatility and intrinsic pump rejection \cite{Schuhmann24}, this provides a distinct advantage over hybrid or monolithic silicon microring SFWM sources.

Finally, combining the above strategies with on-chip electrical pumping \cite{Boitier14,Leger25} and advanced silicon photonic architectures \cite{Silverstone16} will pave the way toward compact and multifunctional quantum photonic chips. Together, these developments will enable robust and scalable platforms for quantum communication, computation, and precision metrology, fully harnessing the potential of hybrid material integration.

\section{Methods} \label{sec:methods}

\subsection*{Device and fabrication}

The AlGaAs structure was grown by molecular beam epitaxy on a GaAs [001] substrate and comprises a 364-nm-thick Al$_{0.45}$Ga$_{0.55}$As core layer, sandwiched between two Bragg mirrors formed by alternating 116-nm-thick Al$_{0.25}$Ga$_{0.75}$As and 280-nm-thick Al$_{0.80}$Ga$_{0.20}$As layers. The bottom mirror consists of six pairs, whereas the top mirror includes two pairs. This asymmetric design provides sufficient confinement for the Bragg mode without compromising the nonlinear conversion efficiency, while simultaneously enhancing the coupling strength to the silicon photonic circuit. The latter is achieved by reducing the separation between the Si waveguide and the AlGaAs core when the epitaxial structure is inverted for heterogeneous integration. Adhesive bonding is performed using benzocyclobutene (BCB), a low-loss transparent polymer, with a calibrated thickness ensuring a final inter-waveguide gap of 40 nm. The structure is optically isolated from the 700-µm-thick silicon substrate by a 1-µm-thick SiO$_2$ layer. For reference, straight AlGaAs and silicon waveguides are fabricated on the same chip, enabling independent characterization. The measured linear propagation losses are approximately 1 cm$^{-1}$ for the AlGaAs waveguides and 3 cm$^{-1}$ for the silicon ones, for both TE and TM polarizations. The AlGaAs generation region -- 2 mm long and 4 µm wide -- is designed to optimize the trade-off between propagation loss and nonlinear conversion efficiency.

\section*{Acknowledgments}

We thank E. Descamps for insightful discussions, J.-R. Coudevylle, E. Herth and P.Filloux for their assistance in the cleanroom and C. Sciancalepore (SOITEC) for providing the SOI wafers.
J. Schuhmann and L. Lazzari acknowledge support through a CIFRE PhD fellowship in collaboration with STMicroelectronics.
This work was supported by the French RENATECH Network, by the Plan France 2030 through projects ANR-22-PETQ-0006 and ANR-22-PETQ-0011, and by the Paris Île-de-France Region within the framework of DIM SIRTEQ through the projects Paris QCI and STARSHIP. It also received funding from the European Union’s Horizon Europe research and innovation programme under the project Quantum Security Networks Partnership (QSNP, Grant Agreement No. 101114043).

%*******************************************
% Bibliography
%********************************************

\bibliography{Biblio.bib}

%********************************************
% Supplementary Information
%********************************************

%\newpage

%\clearpage
%\onecolumngrid

%\setcounter{equation}{0}
%\setcounter{figure}{0}

%\renewcommand{\theequation}{S\arabic{equation}}
%\renewcommand{\thefigure}{S\arabic{figure}}

%\section{Supplementary Material}

% write here the supplementary material, if any

\end{document}